# FORMATION OF THE GALAXY

Sidney van den Bergh

Dominion Astrophysical Observatory,

National Research Council of Canada,

5071 West Saanich Road,

Victoria, British Columbia,

V8X 4M6, Canada

Electronic mail:  vandenbergh@dao.nrc.ca

## ABSTRACT

Current ideas on the formation of the Galaxy are reviewed.  Many of the observed characteristics of our Milky Way System are consistent with a scenario in which the Galaxy formed inside out, with the inner part of it evolving by rapid collapse of a single protogalaxy, while the outer halo was accreted over an extended period.  A number of possible problems with this "Standard Model" are discussed and summarized in § 5.

## 1.    INTRODUCTION

Observational evidence on the formation and early evolutionary history of galaxies may be derived from images and spectra of very distant galaxies (e.g. Cowie, Hu & Songaila 1995, Abraham et al. 1996, van den Bergh et al. 1996) that are viewed at large look-back times.  Alternatively, one can study the fossil evidence for Galactic evolution that is provided by the chemical abundances and kinematics of stars belonging to the oldest population components of the Milky Way System.  More than three decades ago, Eggen, Lynden-Bell & Sandage (1962, ELS) used such observations of the metallicities and orbit shapes of high-velocity stars to conclude that the Milky Way System formed by rapid collapse of a single massive protogalaxy.  More recently, Searle (1977) showed that, contrary to the predictions of the ELS model, globular clusters in the outer halo of the Galaxy did <u>not</u> exhibit a radial abundance gradient.  This conclusion is strengthened by observations of RR Lyrae field stars (Suntzeff, Kinman & Kraft 1991) which show no radial abundance gradient for $R_{GC} > 10$ kpc.  Searle's conclusion, and the remark by Toomre (1977) that "it seems almost inconceivable that there wasn't a great deal of merging of sizable bits and pieces (including quite a few lesser galaxies) early in the career of every major galaxy", led Searle & Zinn (1978, SZ) to propose a scenario for the formation of the Milky Way System in which transient protogalactic fragments "continued to fall into dynamical equilibrium with the Galaxy for some time after the collapse had been completed". Since this model was proposed in the late 1970's, various pieces of evidence have emerged which strengthen the SZ scenario, while others appear difficult to reconcile with it. Some of these problems will be discussed in more detail in this review.

## 2.    FORMATION OF THE GALACTIC HALO

### 2.1    Capture and infall

In situ measurements of the radial velocities of halo stars appear to indicate (Majewski, Hawley & Munn 1996, Majewski, Munn & Hawley 1996) that the Galactic halo may not presently be well-mixed dynamically. This suggests that we are now observing the signature of past infall of the "sizable bits and pieces" envisioned by Toomre (1977). Prima facie evidence for such infall is provided by the recently discovered Sagittarius dwarf galaxy (Ibata, Gilmore & Irwin 1994), which appears to be merging with the Milky Way System. Debris from tidal stripping of the Sagittarius dwarf (and of other dwarf companions of the Galaxy that may once have existed) will result in the formation of moving groups in the halo (Johnston, Spergel & Hernquist 1995). The present absence of any other dwarf companions to the Galaxy with $R_{GC} < 50$ kpc might be due to their removal by capture via dynamical friction (van den Bergh 1994a).

Bellazzinni, Fusi Pecci & Ferraro (1996) have recently noticed a curious relationship between the surface brightnesses and Galactocentric distances of the dwarf spheroidal companions to the Galaxy. They find that the lowest surface brightness (and, hence, the least stable) dSph galaxies are located closest to the Galactic center. Possibly, this result indicates that many low surface brightness dSph galactics remain to be discovered at large distances from the Galaxy. [The faint UMi system would, for example, have been difficult to discover if its red giants had been fainter than the R $\simeq$20 plate limit of the Palomar Sky Survey]. The absence of any dSph galaxies with high surface brightnesses at R < 100 kpc, which is noted by Bellazzini et al., is not unexpected because only three such objects are known in the much larger volume with $100 < R_{GC}$ (kpc) < 300. If the luminosity function of the Local Group does indeed contain more faint objects than presently believed, then the apparent discrepancy between the shallow slope of the Local Group luminosity function, and the steep slope of the luminosity functions of rich clusters (Driver et al. 1994, Bernstein et al. 1995), might be removed.

Van den Bergh (1993) found that 11 Galactic globulars on retrograde orbits have < [Fe/H] > = -1.59 ± 0.07, which is significantly higher than the value < [Fe/H] > = -1.86 ± 0.08 that Suntzeff (1992) derived for 13 true globular clusters in the Large Magellanic Cloud. Since mean cluster metallicity increases with parent galaxy metallicity (van den Bergh 1975), this might be taken to suggest that Galactic globular clusters on retrograde orbits originated in an ancestral galaxy which was more massive than the LMC. Alternatively (and perhaps more plausibly) the fact that the clusters on retrograde orbits have a mean metallicity $< [Fe/H] > = -1.59 \pm 0.07$, which is similar to those of halo clusters on direct orbits (for which $< [Fe/H] > = -1.65 \pm 0.11$) might be understood by assuming that most halo clusters formed in a single highly turbulent protogalaxy.

A Kolmogorov-Smirnov test shows no statistically significant difference in the distribution of [Fe/H] values for 12 halo clusters in prograde orbits and for 11 halo clusters that are in retrograde orbits (van den Bergh 1993). Nevertheless, it is of interest to note that six out of 11 halo clusters in retrograde orbits have metallicities in the narrow range $-1.59 \leq [Fe/H] \leq -1.51$. In fact, all three of the clusters with the most extreme retrograde motions [i.e. those designated R! by van den Bergh (1993)] have metallicities that fall in this narrow range. This is a point that had previously also been noted by Rodgers & Paltoglou (1984). However, the fact that the globular clusters in the LMC have a large range in [Fe/H] values (Suntzeff 1992) shows that the small [Fe/H] range of most Galactic globular clusters in retrograde orbits does <u>not</u> constitute evidence for origin in a single captured ancestral object.  The reason for the small metallicity range of the majority of the Galactic globular clusters on retrograde orbits remains a mystery.

      An argument against the hypothesis that the Galaxy has experienced early mergers with massive ancestral objects similar to the Large Magellanic Cloud is provided by the observation that the globular clusters in the LMC do not exhibit the Oosterhoff (1939) dichotomy of the mean periods of RR Lyrae variables.  A merger with an object similar to the Large Cloud would be expected to have left clusters with intermediate periods, such as that observed in the LMC clusters NGC 1466 ( $<P_{ab}> = 0.59$ days) and NGC 1841 ( $<P_{ab}> = 0.59$ days) in the Galactic halo.  Furthermore, a relatively recent merger with an object resembling the Large Magellanic Cloud or Small Magellanic Cloud would have left many massive intermediate-age clusters in the halo of the Galaxy.  In fact, only half a dozen such objects are known; two of these  (Arp 2 and Terzan 7) (Buonanno et al. 1994) appear to be associated with the Sagittarius dwarf.  This suggests that mergers with rather massive dwarfs resembling the Sagittarius dSph galaxy were probably not <u>very</u> frequent during most of the history of the Galaxy.  This makes it unlikely that such mergers provided a dominant contribution to the population of the Galactic halo.

Minniti, Meylan & Kissler-Pastig (1996) have argued that the cluster Terzan 7, for which Da Costa & Armandroff (1995) find that [Fe/H] = -0.36, is too metal-rich to be physically associated with the Sagittarius dwarf galaxy. However, the same evidence could also be invoked to argue that Ter 7 cannot be a member of the outer halo of the Galaxy!

The youngest known cluster in the Galactic halo is Ruprecht 106 (Kaluzny, Krzeminski & Mazur 1995) which appears to have an age of only 9.3 Gyr (Richer et al 1996). It has been suggested by Lin & Richer (1992) that such relatively young clusters in the Galactic halo might have been tidally captured from the Magellanic Clouds. A possible problem with this hypothesis is that Rup 106 has a half-light radius $r_h$ = 1.10: pc (Harris 1996a), which is less than half that of any other Magellanic Cloud globular. Furthermore, the age of Rup 106 falls near the center of the quiescent period, from 12 Gyr to 6 Gyr ago, during which the Large Cloud does not seem to have produced any cluster. The second youngest known halo cluster is Arp 2, to which Richer et al. assign an age of 10.3 Gyr. This object is probably associated with the Sagittarius dwarf, which precludes it having been detached from either of the Magellanic Clouds. Finally, the third youngest known halo clusters is Palomar 12, to which Richer et al. (1996) assign an age of 10.5 Gyr. Van den Bergh (1994b) gave a number of reasons why this object is unlikely to have been torn from the Magellanic Clouds. In particular, he noted that, at [Fe/H] = -1.14, Pal 12 falls well outside the metallicity range covered by other LMC globular clusters (Suntzeff 1992).

## 2.2    Destruction of globular clusters

Table 1 shows a comparison between the luminosity of all halo stars and the total luminosity of all true globular clusters in the Galaxy, the Large Magellanic Cloud, the Fornax dwarf spheroidal, and the Virgo giant elliptical M49. Perhaps surprisingly, the data in this table show that globular clusters account for rather similar fractions of the total luminosity of the spheroidal populations of these galaxies. Cluster destruction by disk and bulge shocking will be of only negligible importance in the Fornax dwarf galaxy. The close coincidence between the fraction of all halo Population II light that is in the form of globular clusters in Fornax and in the Milky Way System therefor provides weak evidence against the hypothesis (Fall & Rees 1977) that the present Galactic globular cluster system represents only a faint shadow of its former self. This is consistent with calculations by Hut & Djorgovski (1992) which appear to indicate that only 3.6 ± 2.2% of the total Galactic globular cluster population is presently being destroyed per gigayear.

## 2.3    Evolution of the Galactic halo

Hartwick (1978) has argued that the Galactic halo consists of a flattened inner component with c/a ~ 0.6 and a vertical scale-height of 1.6 kpc (which is dominant near the Sun's position), and a more nearly spherical outer component. A somewhat more complex model for the halo is advocated by Norris (1994). According to Norris the Galactic halo consists of (1) a dynamically hot, non-rotating spherical, metal-poor component, within which (2) a somewhat metal-richer ( [Fe/H] > -1.5 ) rotating thick disk is embedded.  Superimposed on this structure is (3) a constituent which was accreted à la Searle & Zinn (1978).  For a more detailed discussion of possible sub-structures in the Galactic halo the reader is referred to the review by Majewski (1993).

Perhaps the strongest evidence against the hypothesis that the outer halo of the Galaxy was mainly formed by capture of "transient protogalactic fragments" (Searle & Zinn 1978) is provided by the observation that the half-light radii $r_h$ of halo globular clusters grow with increasing Galactocentric distance $R_{GC}$.  For clusters with $R_{GC} > 20$ kpc, van den Bergh (1995) finds a rank correlation coefficient $\rho$ ($r_h$ , $R_{GC}$ ) = +0.61 ± 0.18.  One of the reasons for the existence of this correlation is that the Galactic halo contains few compact globular clusters; even though such objects would have survived destruction much more easily than the distended globulars that actually populate the outer halo of the Galaxy.

A Kolmogorov-Smirnov test shows no significant difference between the frequency distribution of half-light radii $r_h$ of globular clusters in the LMC (van den Bergh 1994b) and in the outer ( $R_{GC} > 10$ kpc) halo of the Galaxy (Djorgovski 1993).  However, a comparison between the metallicity distributions of LMC clusters (Suntzeff 1992) and of that for Galactic globulars with $R_{GC} > 10$ kpc shows that the LMC clusters are, on average, more metal deficient than those in the outer halo.  A K-S test shows only a 7% probability that the LMC and Galactic halo clusters could have been drawn from the same parent population of [Fe/H] values.  It is tentatively concluded that capture of a few LMC-like objects is unlikely to have produced a globular cluster population resembling that presently observed in the outer Galactic halo.

In Fig. 1, the half-light radii of the globular clusters in the Fornax dSph galaxy (van den Bergh 1994b) are plotted as arrows.  Note that the radii of these clusters are, on average, smaller than those of globular clusters in the outer halo of the Galaxy.  The paucity of compact clusters, like those in Fornax, in the outer halo of our Milky Way System militates strongly against the hypothesis that the outer halo of the Galaxy was mainly assembled from the debris of Fornax-like dwarf galaxies. Note in particular that all five Galactic globular clusters with $R_{GC}$ > 40 kpc have half-light radii $r_h$ greater than that of the largest Fornax globular.

In Fig. 2, the half-light radii $r_h$ of Galactic globular clusters are plotted versus their perigalactic distances P (van den Bergh 1995). Also shown are the clusters in Fornax plotted at the Fornax perigalactic distance of $100 \pm 40$ kpc (Hodge & Michie 1970). The figure shows that <u>the Fornax clusters are smaller than Galactic globular clusters at similar perigalactic distances</u>. The Sagittarius dSph galaxy, which appears to contain four globular clusters, does not throw much light on the origin of the halo. The cluster NGC 6715 (M54) has $r_h = 2.8$ pc, which is typical of globular clusters in the main body of the Galaxy. On the other hand the cluster Arp 2 has $r_h = 13.2$ pc, which would place it firmly among the outer Galactic outer halo. Finally, Terzan 7 and Terzan 8 have radii of intermediate size.

The half-light radii of globular clusters in the Galactic halo increase with Galactocentric distance $R_{GC}$ and with perigalactic distance P (see Figs. 1 and 2), whereas metallicity [Fe/H] does not (Searle, 1977, van den Bergh 1995). This suggests that <u>the sizes of globular clusters were set by global parameters, whereas</u>

the metallicities of individual clusters were determined by local enrichment events. For very metal-poor halo stars, it is now possible (Sneden et al. 1996) to see the signatures of individual supernova enrichment events.

Lee (1993) and van den Bergh (1993) have independently pointed out that Galactic halo clusters can be divided into two populations on the basis of their metallicities and horizontal branch gradients $C \equiv (B - R) / (B + V + R)$. Fig. 3 shows that all clusters with $R_{GC} < 8.5$ kpc appear to lie on (or close to) a single C versus [Fe/H] relation. Van den Bergh (1993) has assigned all globular clusters of this type to his $\beta$ Population. On the other hand, most halo clusters with $R_{GC} > 8.5$ kpc are seen to fall below (or to the left) of this fiducial line. Van den Bergh (1993) has assigned such clusters to his $\alpha$ Population. Six out of nine clusters (67%) of the $\alpha$ Population, for which orbital data are available, appear to be in retrograde orbits, compared to 2 out of 11 (18%) for clusters of the $\beta$ Population. This suggests that the $\alpha$ Population may, at least in part, have formed in infalling fragments that were subsequently captured by the Galaxy.

Van den Bergh (1993), Lee, Demarque & Zinn (1994) and Da Costa (1994) have argued that the young ($\alpha$) Population of globulars represents objects that were associated with infalling "bits and pieces", whereas the old ($\beta$) Population of clusters belongs to an older protogalactic structure that collapsed <u>à la</u> Eggen, Lynden-Bell & Sandage (1962). A scenario, such as that outlined above, would be consistent with the observation (Da Costa 1994) that clusters of the old halo appear, in the mean, to have direct motion ($V_{rot}$ = +58 ± 24 km s$^{-1}$ ), whereas rotation of the younger halo ($V_{rot}$ = -45 ± 81 km s$^{-1}$ ) is marginally retrograde.

Figure 4 shows a plot of cluster half-light radius $r_h$ versus perigalactic distance P for members of the α Population. Such a correlation would not be expected if clusters of the α Population had been captured at random by the protogalaxy.

A recent compilation of the most accurate ages of individual globular clusters by Richer et al. (1996) shows that eight clusters of the α Population have a mean age < T > = 13.8 ± 0.5 Gyr, while 11 clusters of the β Population are found to have < T > = 15.0 ± 0.4, i.e. the α clusters are, on average, younger than those belonging to the β Population. However, there appears to be a large intrinsic dispersion in the relation between globular cluster age and its distance from the fiducial relation shown in Fig. 3. This suggests (Fusi Pecci et al. 1996) that "second parameter" effects may be a function of both age and of some other, as yet unidentified, factor.

Figure 5 shows plots of $M_V$ versus $R_{GC}$ for globular clusters with red [C $\equiv$ (B-R)/(B + V + R) < -0.80], and with blue and intermediate-color [-0.80 $\leq$ C $\leq$ +1.00] horizontal branches, respectively. Intercomparison of the two panels of this figure shows a significant difference between these two types of clusters in the outer halo ($R_{GC}$ > 10 kpc) of the Galaxy; but no obvious difference in the inner halo ($R_{GC}$ $\leq$ 10 kpc). For clusters with red horizontal branches that are located in the outer halo < $M_V$ > = -4.82, which is almost ten times less luminous than the value < $M_V$ > = -7.30, that is found for outer halo clusters with blue and intermediate-color horizontal branches. A Kolmogorov-Smirnov test shows that there is only a 0.1% chance that the red and blue HB clusters were drawn from the same parent population.

At a given metallicity level globular clusters with red horizontal branches are believed to be younger than ones that have blue horizontal branches (Rood & Iben 1968, Rood 1973, but see Richer et al. (1996)). So the observed effect <u>might</u> be due to a decrease in the luminosity with which clusters are formed over time in the outer halo. Alternatively, this luminosity difference could be related to the fact that red horizontal branch clusters in the outer halo are, in the mean, more metal rich (< [Fe/H] > = -1.32) than are clusters with bluer horizontal branches (< [Fe/H] > = -1.70). Perhaps second generation globular clusters that formed in the outer Galactic halo were, on average, both less luminous <u>and</u> slightly more metal-rich than those formed earlier. However, the low metallicity ([Fe/H] = -1.69) of Ruprecht 106, which is the youngest known halo globular cluster (Kaluzny, Krzeminski & Mazur 1995), would appear to militate against such a simple scenario.

Most dwarf spheroidal galaxies formed stars for a very extended period of time. Such dSph galaxies are observed to contain large numbers of carbon stars. Capture of these dSph galaxies would, therefore, be expected to enrich the Galactic halo in carbon stars. From the (uncertain!) estimates of the number of C stars in the halo, van den Bergh (1994b) estimated that no more than ~40% of the Galactic halo could have been produced by disintegration of Fornax-like dwarf spheroidal galaxies. It should, of course, be emphasized that capture and destruction of dwarf spheroidals that took $\geq 10$ Gyr ago would not have contributed to the carbon star population of the Galactic halo. A much stronger constraint on such captures is set by observations of young blue stars in the Galactic halo (Unavane, Wyse & Gilmore 1996). Such blue objects with B - V $\leq$ 0.4 are younger than the vast majority of halo main sequence stars. Unavane et al. conclude that only ~1% of the halo could have been formed by accretion of dSph galaxies, like the Carina system, that contain a significant intermediate-age population. Preston, Beers & Shectman (1994) use their observations of metal-poor blue horizontal branch stars to conclude that the accreted population is comparable to the total stellar content of all known dwarf spheroidal satellites of the Galaxy.

Some blue stars found in the Galactic halo by Preston, Beers & Shectman (1994) might be the bluest members of a metal-poor intermediate-age population accreted from dwarf spheroidal satellites. Alternatively, a few moderately metal-poor halo A stars (Rodgers, Harding & Sadler 1981, Lance 1988) could perhaps have formed during collisions between metal-poor intergalactic clouds and the metal-rich gas in the disk of the Galaxy. Such metal-poor gas might also have been swept out of one or more dwarf irregular galaxies that collided with gas in the Galactic disk (Freeman 1996).

If the Galactic globular cluster system did, as proposed by Lee (1993), van den Bergh (1993) and Zinn (1993), grow inside out, then the spheroidal cluster component of the Galaxy is presently larger than it was in the past. This contrasts with the situation for the system of open clusters which appears to have been more extended in the past than it is at the present epoch (Friel 1995, Hufnagel 1995). The region with $R_{GC} > 12$ kpc is found to have produced numerous clusters with ages in the range of 2 Gyr - 8 Gyr.  However, no clusters with $R_{GC} >$ 12 kpc seem to have formed during the last 2 Gyr.

Recent HST observations by Richer et al. (1996) appear to show that the outer halo clusters NGC 2419 and Pal. 3, situated at $R_{GC} \sim 100$ kpc, have ages quite similar to those of globulars with $R_{GC} < 10$ kpc.  This has led Harris (1996) to conclude that a "coherent event" must have occurred in the Galactic halo $\sim 15$ Gyr ago.  In this connection, it is of interest to note that the oldest globular clusters in the LMC appear to have ages that are similar to those of the oldest metal-poor globular clusters in the Galactic halo (Brocato et al. 1996).  Taken at face value, these results appear to suggest that the first burst of cluster formation took place almost simultaneously throughout the Galactic halo and in the satellites of the Galaxy.

## 3.  FORMATION OF THE GALACTIC BULGE

The dense absorbing clouds in the direction of the Galactic center render the nuclear bulge of the Milky Way System almost invisible at visual wavelengths.  However, it shows up prominently as a centrally peaked $20° \times 15°$ ($2.8 \times 2.1$ kpc) concentration of IRAS sources at a wavelength of 12 μm (Habing et al. 1985).  Most of these infrared sources are possibly dust-embedded late M giant stars.  High metallicity increases the fraction of giants that become <u>very</u> cool M stars.  The existence of a strong radial metallicity gradient in the inner Galaxy (Terndrup 1988, Frogel et al. 1990, Minniti et al. 1995)  will, therefore, enhance the frequency of IR sources close to the Galactic nucleus.  The distribution of very late M giants, therefore, presents a somewhat biased picture (King 1993) of the distribution of stars in the nuclear bulge of the Galaxy.

An initially steep metallicity gradient in the nuclear bulge will be flattened (Friedli, Benz & Kennicutt 1994) by a central bar (Blitz & Spergel 1991). However, this trend might be partly compensated for by metals produced during ongoing starbursts.  Such star formation presently takes place in the thin, fast rotating, nuclear disk (Dejonghe 1993) which has $R_{GC} < 150$ pc.

Baade's (1951) discovery of large numbers of RR Lyrae stars in the bulge of the Galaxy at first appeared to confirm the hypothesis (Baade 1944) that the nuclear bulge consisted of metal-poor stars of Population II.  However, Morgan (1959) subsequently demonstrated that the <u>dominant</u> population of the Galactic nuclear bulge consists of strong-lined metal-rich stars.  Radial velocity observations of RR Lyrae stars in Baade's Window (Gratton 1987) showed that these objects exhibit a large velocity dispersion ($\sigma \approx 130$ km s$^{-1}$), which clearly marks them as members of the halo population that are just passing through the central region of the Galaxy. A recent study by Minniti (1996) of a bulge field at $\ell = 8°$, b = +7° shows that metal-rich K giants participate in the Galactic rotation ( $< V > = +66 \pm 5$ km s$^{-1}$ , $< \sigma > = 72 \pm 4$ km s$^{-1}$), whereas metal-poor [Fe/H] < -1.5) halo giants ( $< V > = -6 \pm 20$ km s$^{-1}$ , $< \sigma > = 114 \pm 14$ km s$^{-1}$ ) have a large velocity dispersion and do not participate in the rotation of the Bulge. The Bulge has a half-light radius $r_h$ ~200 pc (Frogel et al. 1990). This is an order of magnitude smaller than that of the halo.

From an analysis of the K giants in Baade's window ($\ell = 1°.0$, b = $-3°.9$ Sadler, Rich & Terndrup (1996) find < [Fe/H] > = -0.11 ± 0.04, with more than half of the sample lying in the range -0.4 < [Fe/H] < +0.3. These values probably underestimate the true mean metallicity of bulge stars because (1) the line of sight towards Baade's window intersects the bulge at Z ≈ -0.5 kpc and (b) the sample excludes M giants which will, on average, be more metal-rich than K giants. From studies of the <u>integrated</u> light of stars in Baade's window, Idiart, de Frietas Pacheco & Costa (1996) find [Mg/Fe] = +0.45 in the Galactic nuclear bulge.

The very high metallicities of some bulge stars have recently been confirmed with Keck echelle spectra obtained by Castro et al. (1996). These authors find [Fe/H] = +0.47 ± 0.17 for the star BW IV -167.  This is similar to the value [Fe/H] = +0.46 ± 0.14 obtained for the nearby super metal-rich star μ Leonis.

Minniti (1995) concludes that metal-rich globular clusters with $R_{GC} > 3$ kpc belong to the Thick Disk, but that those having $R_{GC} < 3$ kpc are kinematically associated with the bulge.  The color-magnitude diagrams of these clusters also appear consistent with their assignment to the bulge population. From color-magnitude diagrams obtained with the Hubble Space Telescope (HST), Ortolani et al. (1996) conclude that the ages of the bulge clusters NGC 6528 and NGC 6553 do not differ by more than a few Gyr from that of the Thick Disk cluster 47 Tucanae.   It has, however, been emphasized by Catalan & de Frietas Pacheco (1996) that such small age differences are rendered uncertain by possible differences in helium abundance and in the ratios of elements produced by SNe Ia and SNe II.

From rather noisy color-magnitude diagrams that extend down to the main sequence turnoff in Baade's Window, Terndrup (1988) concluded that the bulk of the stars in the bulge have ages in the range of 11 - 14 Gyr.  Furthermore, he found that the number of objects with ages < 5 Gyr is negligible.

The conclusion that disks and bulges of galaxies belong to distinct population components is supported by observations of M33.  This galaxy has a well-developed halo containing globular clusters (Schommer et al. 1991) and RR Lyrae stars (Pritchet 1988), but appears to have little or no old bulge (Bothun 1991).  In other words, a galaxy can have a halo but no bulge.  However, the central region of M33 does contain evolved stars that are brighter than those in the Milky Way bulge (Minniti, Olszewski & Rieke 1993).  Mighell & Rich (1995) suggest that such stars are associated with a relatively recent burst of star formation that took place well after the oldest stars were formed.

Hartwick (1976) proposed that the Galactic disk was formed by gas ejected from halo stars.  However, this suggestion appears difficult to reconcile with the high specific angular momentum of disk stars.  More recently, Carney, Latham & Laird (1990) and Wyse & Gilmore (1992) have suggested that the Galactic bulge was formed from low angular momentum gas that was left over after most star formations had ended in the halo.  The idea that halo stars were formed from leftover halo gas, which was enriched by SNe II on a short time-scale, appears to be supported by McWilliam & Rich (1994) who find that, compared to disk stars near the sun, [Mg/Fe] and [Ti/Fe] are (as is the case in halo stars) elevated by $\approx$ 0.3 dex.  However, it is not clear why McWilliam &

Rich find [Ca/Fe] and [Si/Fe] in bulge stars to closely follow normal trends for Galactic <u>disk</u> giants. Sadler, Rich & Terndrup (1996) find < [CN/Fe] > to be close to solar in metal-poor stars in the bulge, whereas < [CN/Fe] > = -0.47 ± 0.04 in metal-rich stars. The weak CN lines in bulge stars indicate that these objects differ in some respects from stars in elliptical galaxies. An additional complication (Ratag et al. 1992) is that He and N in bulge planetary nebulae appear to be higher than they are in the Galactic disk. All of these results suggest that the evolutionary history of the Galactic bulge was probably complex. This conclusion is supported by the observation that the central region of the Galaxy presently contains much less gas than would have been ejected by the stars in the Galactic bulge during a Hubble time (van den Bergh 1957). Most of the gas lost by first-generation bulge stars was probably used up to form second-generation stars. Such second-generation stars could have incorporated elements produced on a relatively long time-scale by SNe Ia.

In elliptical galaxies, and in the large bulges of Sa and Sb galaxies that collapsed rapidly, the absence of gas can probably be accounted for (Mathews & Baker 1971) by invoking galactic winds generated via gas heating caused by supernova blast waves. Sofu & Habe (1992) postulate that such bulges are themselves formed by star bursts in gas clouds ejected from the central regions of galaxies. On the other hand, Sellwood (1993) has suggested that nuclear bulges might have formed as the result of bar-like instabilities in disks. However, Minniti (1995) questions whether the steep abundance gradients observed in the nuclear bulges of galaxies could have been formed (or maintained) if disks had first been stirred up by bars. It is still too early (Renzini 1993) to decide if any (or all) of the processes discussed above contributed to the formation of the Galactic bulge. Finally, it is noted that Lee (1992) believes the bulge of the Galaxy to be older than the halo. Renzini & Greggio (1990) have also argued that the bulge formed <u>before</u> the halo because the collapse time scale $\tau = (G\rho)^{-\frac{1}{2}}$ is much longer for the halo than it is for the bulge. So the bulge may be older than the halo, even though it is metal-richer than the halo.

A possible example of an infalling object is the globular cluster NGC 6287 (Stetson & West 1994). This cluster is very metal-poor ( [Fe/H] = -2.05) and is located at only 1.9 kpc (Djorgovski 1993) from the Galactic nucleus. Its small half-light radius of $r_h$ = 1.3 pc suggests that it  might be physically associated with the central region of the Galaxy. However, Stetson & West point out that the relatively high radial velocity (V = -208 km s$^{-1}$ ) of NGC 6287 gives it sufficient kinetic energy to travel out (or fall in from) as far as the Solar circle.

Some insight into the formation of spiral galaxies is provided by the morphology of individual spirals in the Hubble Deep Field (e.g. van den Bergh et al.1996).  A good example is the spiral HDF 2-86.  This object has a nucleus which is slightly orange in color, indicating the presence of some evolved stars.  This nucleus is embedded in a disk (or flattened clustering) of presumably younger blue knots.  This indicates that <u>the nuclear bulge may already be present in a proto-spiral before assemblage of the disk is complete</u>.

## 4.    FORMATION OF THE GALACTIC DISK

### 4.1    Evolution of the Galactic disk

The evolutionary relationships between the Galactic halo, the Thick Disk and the Thin Disk remain a subject of lively controversy.  In particular, it is not yet clear whether the transition from the halo phase of Galactic evolution to the disk phase was continuous, or if there was an extended hiatus (Berman & Suchkov 1991) between the halo and disk stages of evolution.  If there were such a hiatus, then it would no longer be possible to regard the Thick Disk as a structure that formed when pressure support started to build up at the beginning of the dissipational phase of Galactic evolution.  It is not yet clear [see Majewski (1993) for an excellent review] whether the Thick Disk and the Thin Disk represent distinct evolutionary phases, or if there was a gradual transition between the Thick Disk and Thin Disk eras of Galactic evolution.  Observations by Oswalds & Risley (1961) show that short-period Mira variables exhibit halo kinematics, whereas longer period Miras seem to belong to a Thick Disk population.  More detailed studies of Mira variables might, therefore, provide some insight into the nature of the Thick Disk/Thin Disk transition.  Realistic numerical simulations, which take into account energy and angular momentum transfer to the halo (Barnes 1996), seem to show that the Thick Disk probably does <u>not</u> represent Thin Disk material that was heated by capture of an infalling satellite. [If this view is correct, then the existence of thin disks in many spirals no longer places strong constraints (Tóth & Ostriker 1992) on the rate at which disk galaxies capture companions].  The observation by Gratton et al. (1996) that [Mg/Fe] $\approx$ +0.4 in the Thick Disk, but that it decreases by 0.2 dex at the Thick Disk/Thin Disk transition, also militates against the suggestion that the Thick Disk consists of dynamically heated Thin Disk stars.  However, this conclusion depends critically on how the Thick Disk to Thin Disk transition is defined. Observations of a few nearby edge-on spirals (Morrison 1996) may indicate that only galaxies with central bulges exhibit Thick Disks.  The reason for the possible existence of such a relationship between bulges and thick disks is not immediately obvious.  The reality of a metal-poor (and hence very old) population component in the Galactic disk is presently in doubt (Twarog & Anthony-Twarog 1996).

Realistic merger calculations (Barnes 1996) suggest that it may not be possible to account for very metal-poor disk stars by invoking capture (Quinn & Goodman) of, and mergers with, dwarf galaxies.

## 4.2    The open cluster system

Recently, Friel (1995) has studied 74 open clusters with ages larger or equal to that of the Hyades.  For these disk clusters, she finds that:  (1) clusters with $R_{GC} \lesssim 7$ kpc are absent; presumably because they have been destroyed by interactions with giant molecular clouds (van den Bergh & McClure 1980).  (2) Open clusters (and, hence, the thin disk) exhibit a steep metallicity gradient with $< [Fe/H] > \simeq 0.0$ at $R_{GC} = 7$ kpc and $< [Fe/H] > \simeq -0.5$ at $R_{GC} = 12.5$ kpc.  (3) At any value of $R_{GC}$ , open clusters show a range in metallicity of about 0.5 dex.  (4) Over the range $7 \leq R$ (kpc) $\leq 13$ open clusters exhibit no evidence for a dependence of metallicity on age.  Edvardsson et al. (1993) have recently obtained a similar result for element abundances (excluding Ba) in field stars with ages < 10 Gyr.  The most straightforward interpretation of this result is that the heavy elements produced by supernovae are diluted by infall of more-or-less pristine gas into the disk.  Presumably, such infalling gas will, on average, have zero angular momentum.  Such infall will reduce the angular momentum of the Galactic disk and result in its radial contraction.  Possible evidence for such infall is also provided by the observation (Edvardsson et al. 1993) that the scatter in [Si/Fe] is about four times *smaller* than that in [Fe/H]. This is exactly what would be expected if disk stars form from gas that has not been well-mixed after infall of clouds with more-or-less pristine composition. Within the Galactic disk, Edvardsson et al. find that $[\alpha / Fe]$ decreases with increasing $R_{GC}$. This suggests that the rate of star formation declined faster in the inner disk than it did at larger radii.

For a detailed discussion of the chemical and stellar evolution of the Galactic disk the reader is referred to Prantzos & Aubert (1995). However, a problem is that constraints imposed by the disk oxygen abundance are uncertain. This is so because H II regions show a steep [O/H] gradient in the outer disk, whereas observations of B stars appear to exhibit no such gradient.

### 4.3    Age of the Galactic disk

The Thick Disk of the Galaxy is observed to contain many RR Lyrae variables. Since RR Lyrae stars occur in the "young" globular cluster Ruprecht 106 (Kaluzny, Krzeminski & Mazur 1995), to which Richer et al. (1996) assign an age of 9.3 Gyr, it follows that the Thick Disk must have an age of at least 9 Gyr. This conclusion is marginally consistent with the fact that few carbon stars

(most of which are thought to have ages $\leq$ 10 Gyr) appear to have Thick Disk

kinematics. An even greater age of $12 \pm 2$ Gyr has recently been obtained by

Phelps et al. (1996) for the probably open cluster Berkeley 17. If one <u>assumes</u>

that the disk formed after the end of the halo phase of Galactic evolution, then an upper limit on the age of the Galactic disk is set by the halo cluster M92, for which Bolte & Hogan (1995) derive an age of $15.8 \pm 2.1$ Gyr. Other normal halo globular clusters may be younger than M92, but have less well-determined ages.

[Young halo clusters such as Rup 106 and Ter 7 probably had unusual

evolutionary histories (van den Bergh 1996) and may have formed after the thick

disk was assembled]. A weaker, but entirely independent, upper limit to the age

of the Galactic halo is set by the thorium abundance in the ultra metal-poor star

CS 22893-052, which yields an age of $16 \pm 6$ Gyr (Sneden et al. 1996).

However, the work by Phelps, Janes & Montgomery (1994), and of Kaluzny,

Krzeminski & Mazur (1995) shows that there may be some overlap between the ages of the oldest open clusters and those of the youngest globulars. In other words, the disk may have started to form before formation of the halo was completed.

 

If all open clusters are members of the Thin Disk then a lower limit to the age of the Thin Disk is provided by NGC 6791, which is the oldest open cluster with a well-determined age. Garnavich et al. (1993), Kaluzny & Rucinski (1995) and Tripicco et al. (1995) find ages of 7 - 10 Gyr for this object. The motion of this cluster, which lags circular motion by more than 60 km s $^{-1}$ (Scott, Friel & Janes 1995) is, however, somewhat peculiar. An age similar to that of NGC 6791 is obtained from the calculated production ratios of the actinoid pairs $^{235}$U/$^{238}$U and $^{232}$Th/$^{238}$U and their presently observed abundance ratios. From these values Chamcham & Hendry (1996) find that star formation in the solar neighborhood began at least 9 Gyr ago.

An entirely independent lower limit to the age of the Galactic disk is provided by the colors and trigonometric parallax (Ruiz et al. 1995) of the white dwarf ESO 439-26, which is found to have $M_V = +17.6 \pm 0.1$. This low luminosity yields a cooling age of 6 - 7 Gyr. In summary, it appears that all presently available data appear consistent with ages of between 12 Gyr and 15 Gyr for the Thick Disk of the Galaxy. Berkeley 17, the oldest known "open" cluster may, on the basis of its kinematics (Scott, Friel & Janes 1995), be a member of the Thick Disk. The age of this cluster is estimated to be $12 \pm 2$ Gyr, which is consistent with the Thick Disk age limits given above.

## 5. SUMMARY AND CONCLUSIONS

It is now widely believed that the central region of the Milky Way System collapsed from a single protogalaxy à la ELS, while the outer part of the Galaxy is thought to have been assembled by infall (and subsequent capture) of protogalactic fragments, as envisioned by SZ. The observation that 6 out of 9 young globular clusters of $\alpha$-type have retrograde orbits, whereas only 2 out of 11 older $\beta$-type clusters have retrograde orbits, appears consistent with this scenario. In the "Standard Model", the Galaxy formed inside out, with the bulge being old and the outer halo relatively young. However, the following problems are noted with this Standard Model:

(a)     The Standard Model does not account for the observation (see Fig. 1) that the half-light radii $r_h$ of globular clusters grow with increasing Galactocentric distance $R_{GC}$ . An even closer (and as yet unexplained!) relation (Fig. 2) exists between $r_h$ and perigalactic distance P. It is particularly puzzling (see Fig. 4) that young $\alpha$-type globular clusters in the halo, which should mainly be captured objects, appear to exhibit a close correlation between $r_h$ and P.

(b)     The globular clusters associated with the Fornax dwarf spheroidal galaxy are much smaller (see Fig. 2) than the globular clusters in the Galactic halo.  It follows that the halo cannot have been entirely assembled by capture of Fornax-like dSph systems.  The frequency of blue intermediate-age stars in the halo also limits <u>recent</u> accretion of Carina-like dSph systems to ~1% of the total stellar population of the Galactic halo.

(c)    The periods of RR Lyrae stars in Galactic halo globular clusters exhibits a marked Oosterhoff  dichotomy, with mean cluster periods

$< P_{ab} > \simeq 0.55$ days and $< P_{ab} > \simeq 0.65$ days.  No such dichotomy is observed among the RR Lyrae variables in globular clusters associated with the LMC.  This suggests that the Galaxy did not merge with one or more galaxies resembling the Large Cloud during the course of its evolutionary history.  The fact that only about half a dozen Galactic globular clusters are known to have ages as low as ~10 Gyr suggests that the number of mergers with Sagittarius-like dSph galaxies has probably been small.  Weak limits on the number of recent mergers with dSph galaxies can also be set from the frequency of C stars in the Galactic halo.

(d)    Contrary to expectations from the Standard Model, recent HST observations of the clusters NGC 2419 and Pal. 3, at R ~100 kpc, yield ages that are similar to those of typical globular clusters at R < 10 kpc.

This suggests the possibility that a "coherent event" might have produced a burst of cluster formation in the Galaxy ~15 Gyr ago.

(e)     It is not clear (see Fig. 5) why globular clusters in the outer halo, that have red horizontal branches, are an order of magnitude <u>less</u> luminous than those that have bluer horizontal branches.

Taken at face value, some of these results appear to weakly favor a scenario (Sandage 1989) in which mergers represent "noise" that is superposed on an ELS-like collapse model for the Galactic halo.  It should, however, be emphasized that infall might have played a more important role in the evolutionary history of other giant galaxies.  The observation that globular cluster radii $r_h$ correlate with $R_{GC}$ , but that [Fe/H] does not, appears to favor a scenario in which cluster sizes are set by global parameters, whereas their metallicities are determined by local factors.

I thank Drs. Michael Bellazzini, Mike Bolte, Raffaele Gratton, Ron Marzke, Flavio Fusi Pecci, María Teresa Ruiz, Nick Suntzeff and Matt Wood for exchanges of views and useful information.  I also wish to thank a particularly helpful referee.

TABLE 1

Comparison between luminosities of halo Population II and the integrated
luminosity of globular clusters

| Galaxy globulars | $M_V$ (halo) | $M_V$ (globulars) | $\triangle M_V$ | L (globulars) |
|---|---|---|---|---|
| Milky Way | -18.4[a] | -13.0[a] | -5.4 | 0.7% |
| LMC | -15.1[a] | -10.9[a] | -4.2 | 2.1% |
| Fornax | -13.7:[c] | -8.8[b] | -4.9 | 1.1% |
| M49 | -22.9[d] | -17.7 | -5.2[d] | 0.8% |

[a]    Suntzeff (1992)

[b]    van den Bergh (1995)

[c]    Total luminosity

[d]    Harris (1991)

# FIGURE CAPTIONS

Fig. 1  Half-light radii of Galactic globular clusters versus Galactocentric

distance.  The radii of globulars are seen to increase with distance from the Galactic center.  Globular clusters in the Fornax dwarf are shown as horizontal arrows.  Note that the Fornax clusters are more compact than most globular clusters in the outer halo of the Galaxy.

Fig. 2  Half-light radii of Galactic globular clusters versus cluster perigalactic

distances.  The Fornax clusters (open circles) are plotted at the $100 \pm 40$

kpc perigalactic distance of this dwarf spheroidal galaxy.  The Fornax clusters are seen to be smaller than Galactic globulars at such large perigalactic distances.

Fig. 3  Metallicity versus horizontal branch population gradient in the inner (top)

and outer (bottom) regions of the Galactic halo.  Clusters that fall close to the fiducial line are assigned to the β Population.  Clusters below (or to the left of) this line belong to the (possibly younger) α Population.

Fig. 4  Relation between the perigalactic distance P (van den Bergh 1995) and half-light radius $r_h$ (Djorgovski 1993) for young metal-poor halo globular clusters belonging to the $\alpha$ Population (van den Bergh 1993). After excluding the only collapsed-core cluster in the sample (shown as a cross), it is found that the correlation coefficient between $r_h$ and P is r = 0.95 $\pm$ 0.03: . Such a strong correlation between cluster radius and perigalactic distance is not expected for a scenario in which such clusters were captured, more or less at random, from the neighborhood of the Galaxy.

Fig. 5   Integrated magnitudes versus Galactocentric distances for clusters with red horizontal branches having C < -0.80 (bottom) and for clusters with C $\geq$ -0.80 (top). For R > 10 kpc clusters with red horizontal branches are seen to be almost ten times fainter than those with bluer horizontal branches.

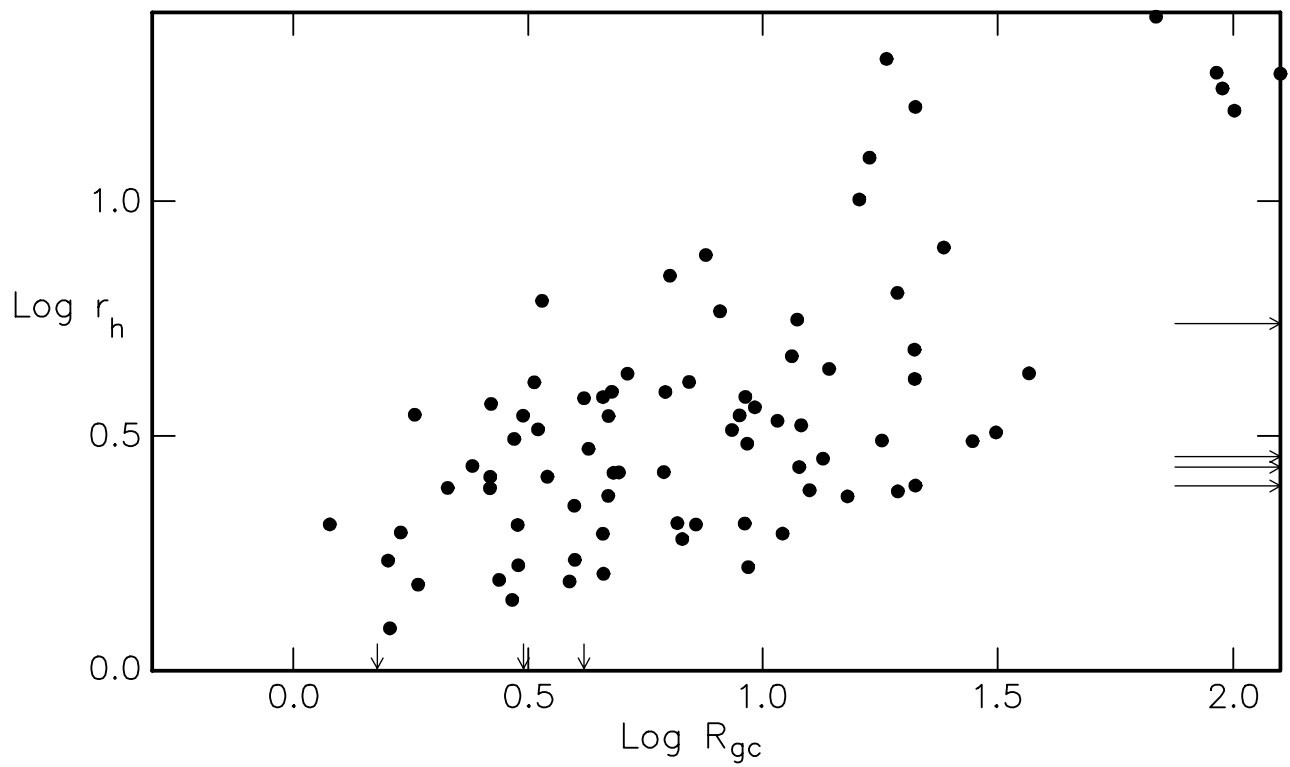

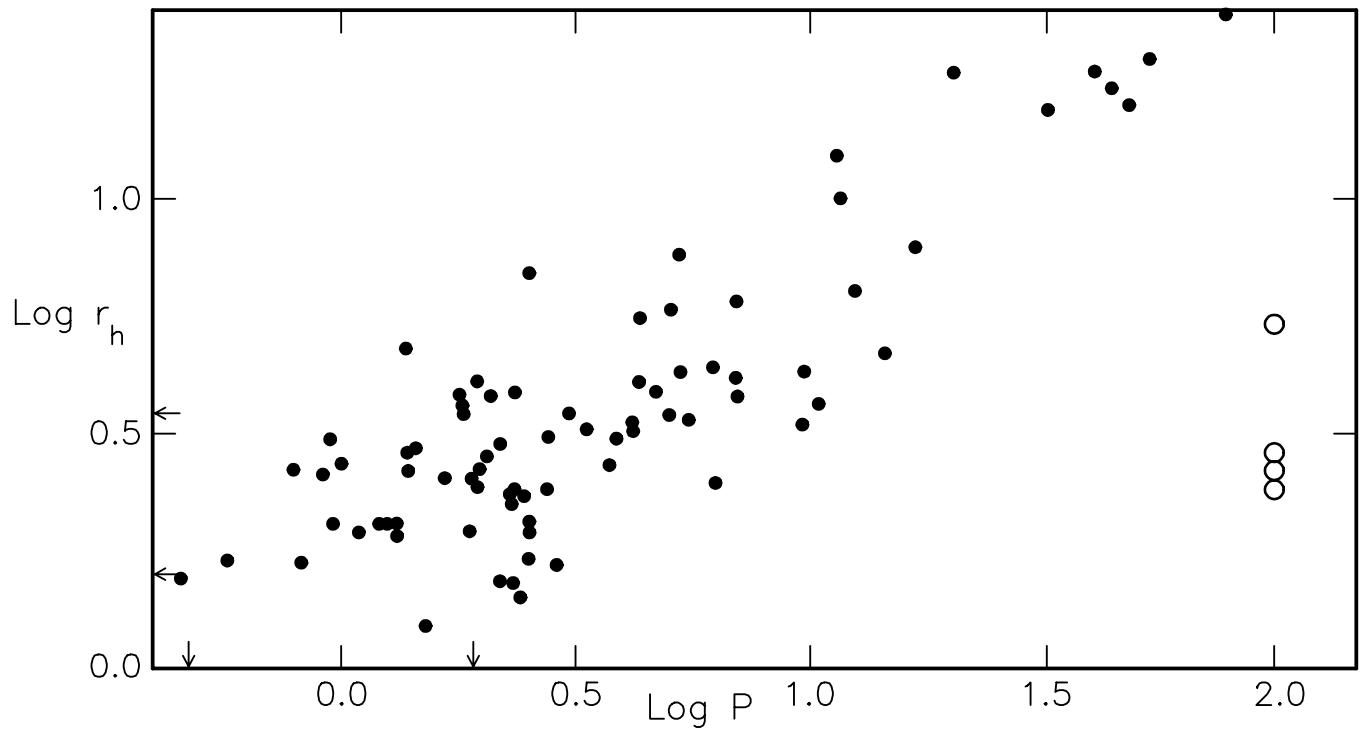

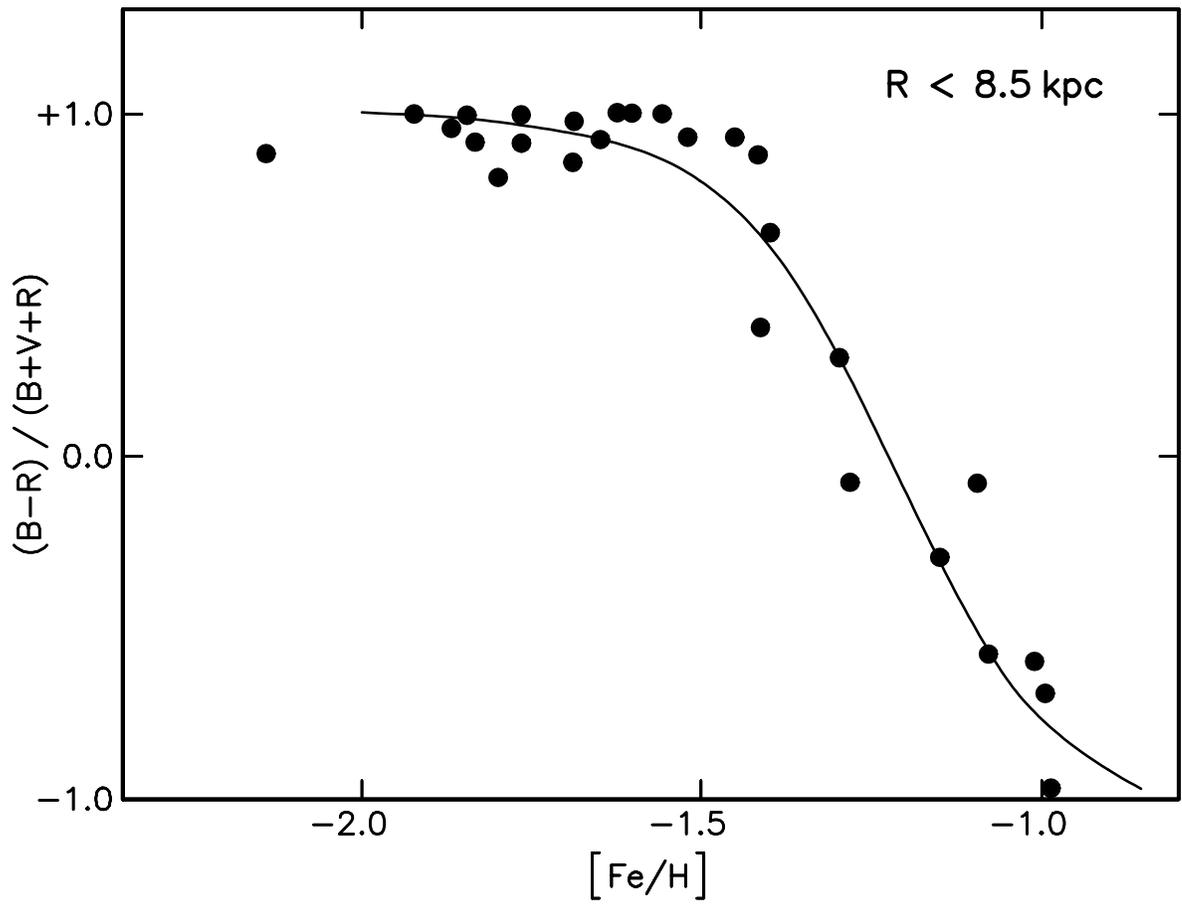

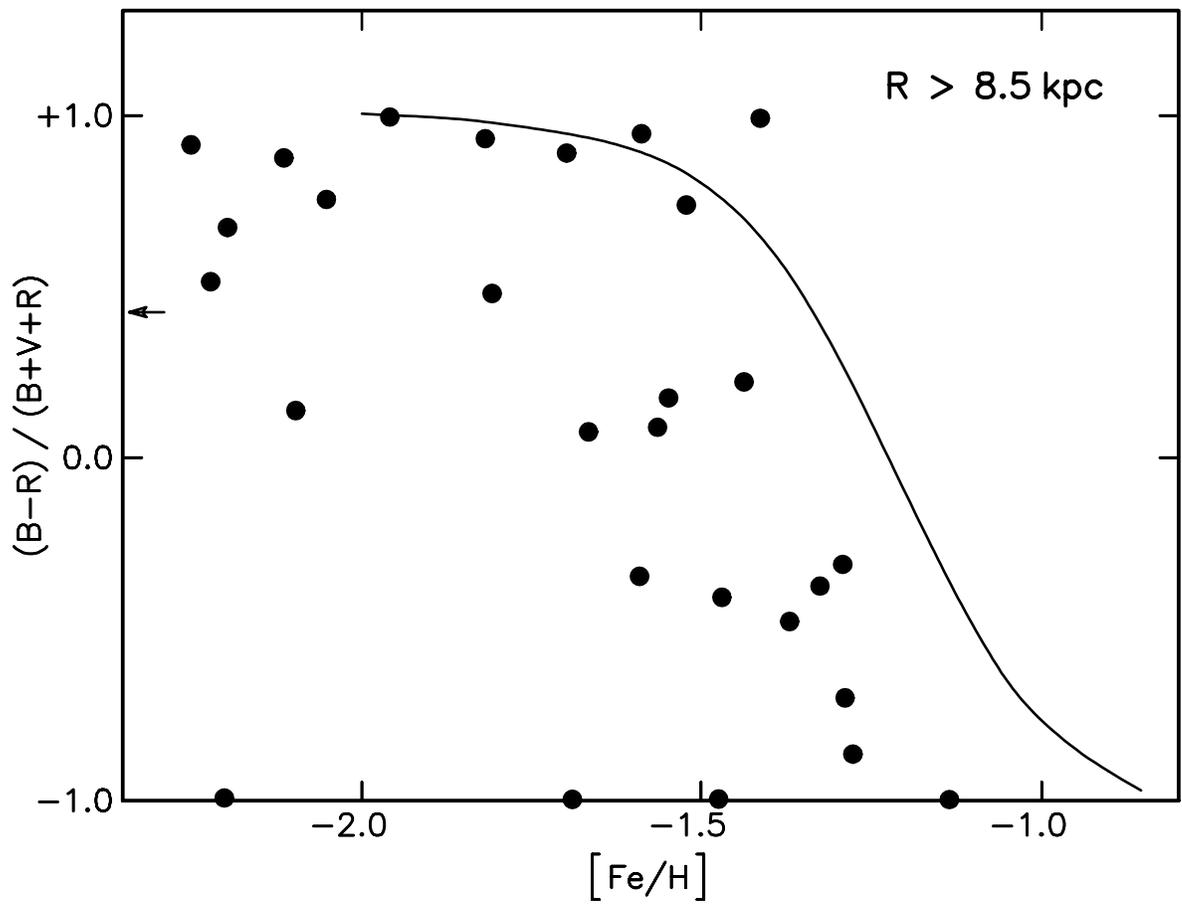

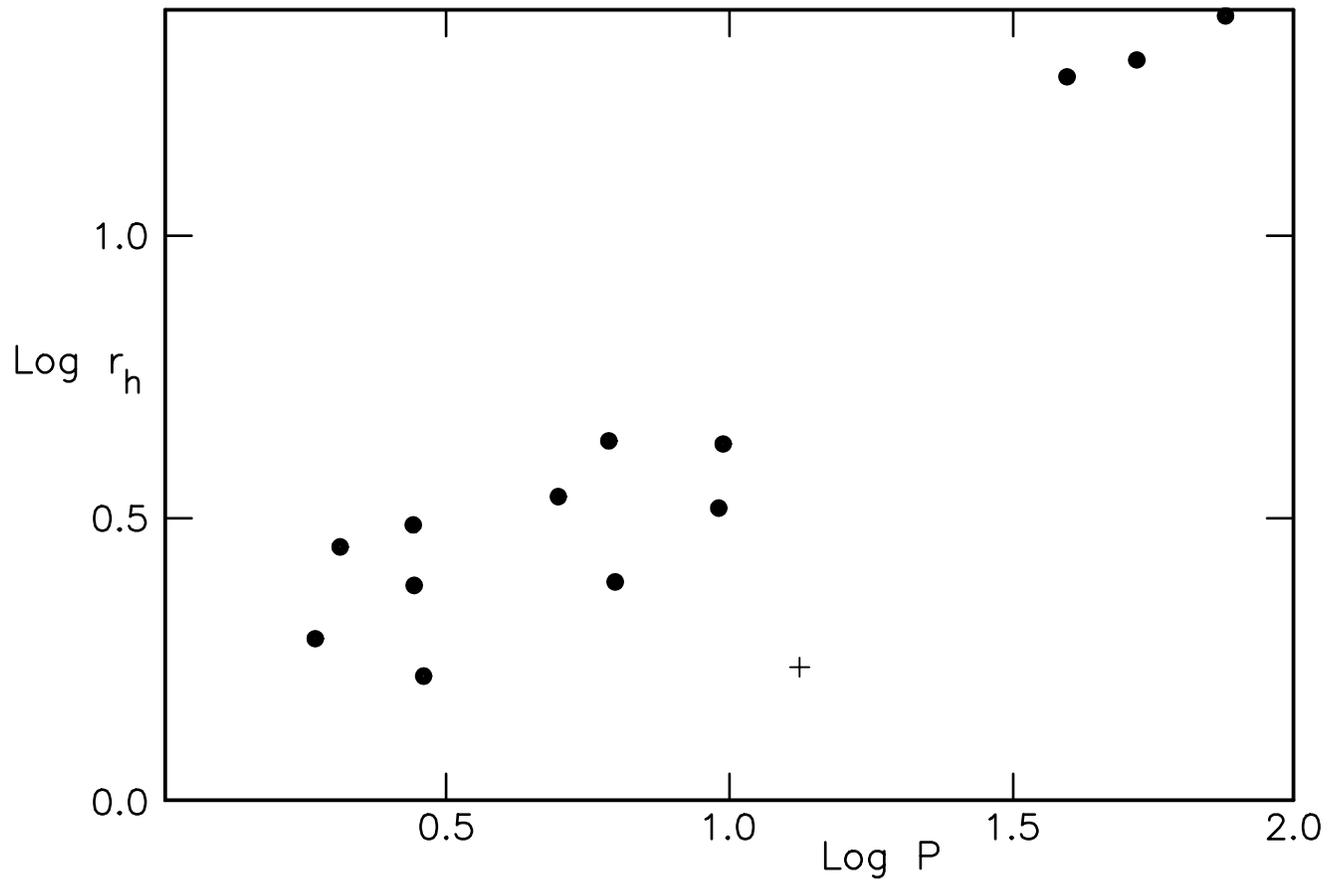

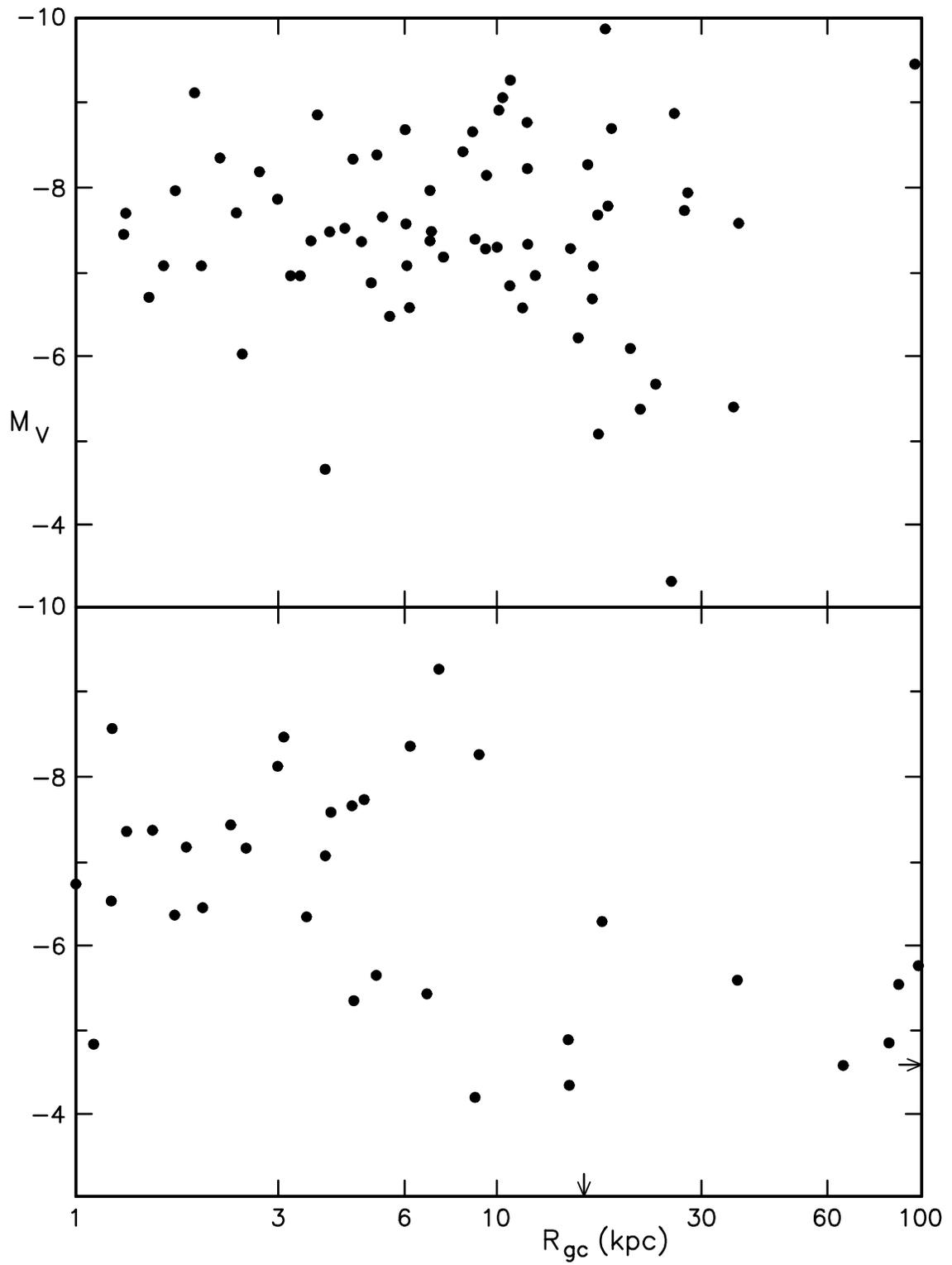